\documentclass[english]{elsarticle}
\usepackage[T1]{fontenc}
\usepackage[utf8]{inputenc}
\usepackage{textcomp}
\usepackage{graphicx}
\usepackage[dvipsnames]{xcolor}
\usepackage{cancel,soul,ulem}

\usepackage{array}
\usepackage{wrapfig}
\usepackage{multirow}
\usepackage{tabularx}
\newcommand{\comment}[1]{}

\newcommand{\microns}{~$\mu$m}

\makeatletter
\@ifundefined{date}{}{\date{}}
\makeatother

\usepackage{babel}
\begin{document}

\begin{frontmatter}{}

\title{Heat flux balance description of unidirectional freezing and melting dynamics on a translational temperature gradient stage} 

\author[huji]{Michael Chasnitsky \corref{cor1}}

\ead{michael.chasnitsky@mail.huji.ac.il}

\author[paris]{Victor Yashunsky \corref{cor1}}

\ead{victoryashunsky@gmail.com}

\author[huji]{Ido Braslavsky\fnref{fn2}}

\address[huji]{The Robert H Smith Faculty of Agriculture, Food and Environment,
The Hebrew University of Jerusalem, Rehovot, Israel }

\address[paris]{Laboratoire Physico-Chimie Curie, Institut Curie, PSL Research University—Sorbonne Universit\'e, UPMC-CNRS—Equipe labellis\'ee Ligue Contre le Cancer, 75005 Paris, France}

\cortext[cor1]{Corresponding author}

\fntext[fn2]{www.agri.huji.ac.il/\textasciitilde braslavs}
\begin{abstract}
Directional solidification occurs in industrial and natural processes,
such as freeze-casting, metal processing, biological cryopreservation
and freezing of soils. Translational temperature gradient stage allows to control the process of directional solidification and to visualise it with optical microscope.
In this stage freezing velocity and temperature gradient are decoupled and are independently controlled. 
Here we study the dynamics of the phase transition interface
in thin water samples using translational temperature gradient stage.
We follow position of the ice-water interface with optical microscopy and compare it to solution of one dimensional Stefan problem in the low velocity limit. 
We find an agreement between experimental observations and theoretical predictions for constant velocity and during acceleration of the ice front. This work presents a practical
framework for analysis and design of experiments on a translational
temperature gradient stage.

\end{abstract}

\end{frontmatter}{}

\section{Introduction}

Understanding solidification and freezing processes is key for material
science \citep{zhang2005aligned,deville2006freezing,glicksman2010principles}
and biology \citep{rubinsky1988mathematical}, such as in the widespread
process of freeze casting or
ice templating material fabrication method, where the growth of ice crystals shapes suspended
colloids. In 1965, \citet{hunt1966temperature} introduced a translational
temperature gradient stage for the microscopic study
of the solidification process in transparent organic compounds, as
a model system for studying metal solidification. In the 1980's, a
similar stage was applied to study cryopreservation of
biological samples and freeze casting studies \citep{brower1981hypothesis,rubinsky1985cryomicroscope,korber1983solute}. 
Recently this setup was used  for 3D imaging of ice growth \cite{dedovets2018temperature,dedovets2018five}, measurement of water freezing  point  depression \cite{PhysRevE.67.061602}, ice-templating of ceramics \cite{stolze2016directional} and for cryopreservation of adherent cell cultures \cite{bahari2018directional}.

As the velocity is controlled externally, the translational temperature gradient stage decouples the freezing velocity and the thermal gradient. 
It also allows to resolve the position and structure of the phase interface at high resolution using microscopy.
Experimental studies executed on the translational temperature gradient stage often assume constant gradient, ignoring the thermal gradient difference in solid and liquid media in the freezing sample \cite{You2015,you2016interfacial,schollick2016segregated}.

Several attempts were done to formulate general description or numerical algorithms to describe the freezing on translational temperature gradient stage. 
For example, uni-dimensional temperature distribution was calculated numerically for broad freezing velocity range \citep{lipp1987investigation,rabin2000reasonable,Hagiwara2012} and an algorithm for calculation of three-dimensional temperature field in case of slow freezing velocities was proposed \citep{Ukpai2019}.
Such general solutions require heavy computations and do not provide a predictable observables that could be directly accessed through experiment.

\begin{figure}
\begin{centering}
\includegraphics[width=12cm]{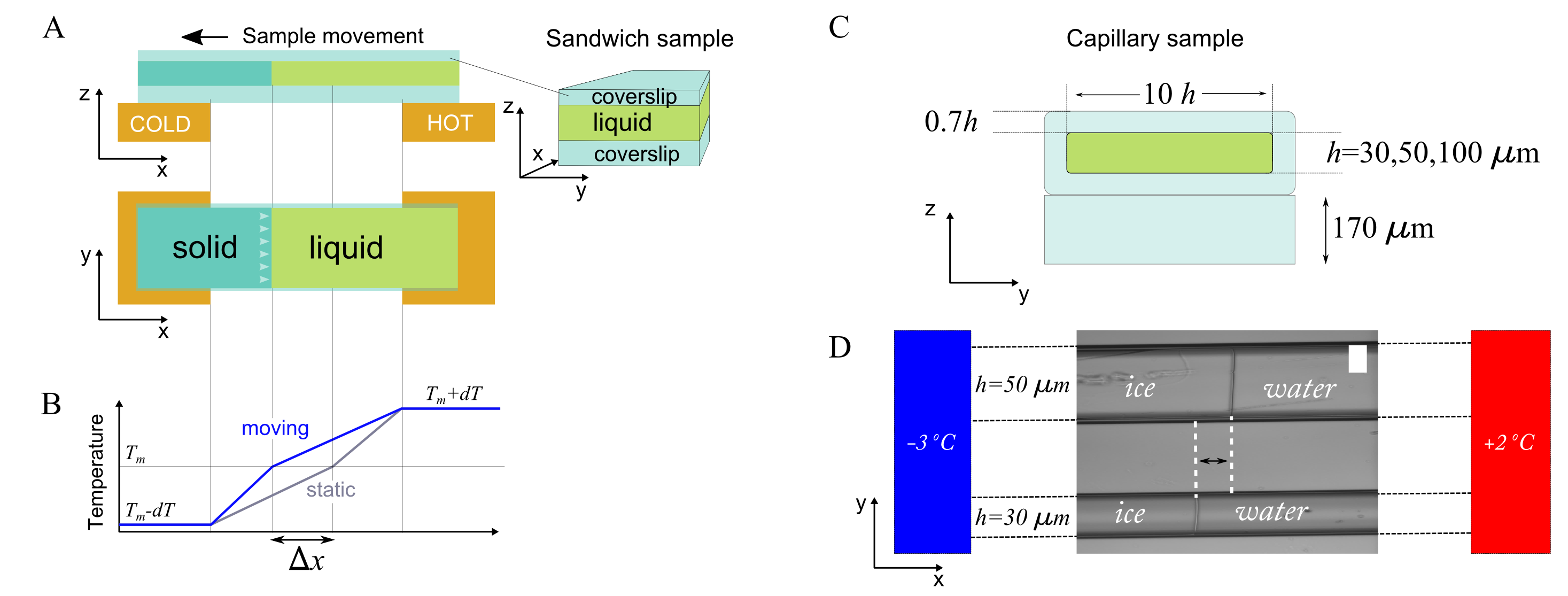}
\par\end{centering}
\caption{\label{fig:stage-setup}Translational directional stage setup scheme.
(A) Sample in between glass coverslips moves on top of thermal blocks.
(B) The temperature of the blocks set above and below the melting
point $\left(T_{m}\pm dT\right)$ producing a temperature gradient
in the gap between the blocks and the resulting solid-liquid interface
in the gap. The interface position is determined by sample velocity
($v_{s}$) along with the cross-section ratio of liquid and glass
in the sample. $\Delta x$ indicates the shift of the interface from
its position in the static sample. (C) A cross-section of a capillary
sample: the capillary was placed on top of a glass coverslip. (D)
The experimental position of the interface differs between the capillaries
when the sample is static and the temperatures of the thermal blocks
are 2 degrees above the melting point for the hot block and 3 degrees
below for the cold one. The scale bar is 200{\microns} in
height.}
\end{figure}

In this work, we introduce a theoretical framework that allows the prediction of the ice front position during freezing and melting of pure water  (and other pure substances) at low velocities for design and analysis of experiments on the translational temperature gradient stage.
We compare the experimental results to the theoretical predictions studying different sample geometries and range of ice front velocities during freezing and melting. Specifically, we investigate the propagation of the ice front at constant velocity and in the course of acceleration.

\section{Material and methods }
\subsection{Physical description of translational gradient setup}

The translational gradient stage consists of two independently controlled thermal blocks separated by thermally non conductive medium (Fig. \ref{fig:stage-setup}A). 
The temperature of the thermal blocks set above and below the melting point ($T_m$), namely temperature boundary conditions are $T(x=0)=T_{m}-dT_{c}$ and $T(x=L)=T_{m}+dT_{h}$ for the cold and the hot block respectively. 
The sample placed on the thermal blocks and is subjected to the temperature gradient as ice front appears in the gap between the thermal blocks (Fig. \ref{fig:stage-setup}B). 

\subsection{Stage}

Custom microscope translational
temperature gradient stage was build as previously described in details \citep{bahari2018directional}.
The setup shown schematically in Figure \ref{fig:stage-setup}A. It consists of two independently-controlled Cu blocks at temperatures
($T_{m}-\Delta T_{c},\,T_{m}+\Delta T_{h}$) separated by a 2.6 mm
air gap. The sample is mounted on top of the thermal blocks and
moved laterally by the linear actuator (CONEX-TRA12CC, Newport, USA).

The temperatures at the edges of the thermal blocks slightly differ from the set temperatures by temperature controller (PRO8000, ThorLabs, USA). To compensate for this effect the temperature at the edges approximated as ($T_{m}-dT_{c},\,T_{m}+dT_{h}$). The prefactor $dT_{c/h}=0.7\cdot\Delta T_{c/h}$ was estimated from the simulations performed with COMSOL software  for our translational gradient stage \citep{bahari2018directional}.

The entire setup controlled trough in house written Labveiw  software including video acquisition, temperature control, sample position and velocity.

\comment{
($T_{m}-\Delta T_{c},\,T_{m}+\Delta T_{h}$) differ from the temperatures
at the edges of the gap ($T_{m}-dT_{c},\,T_{m}+dT_{h}$), which are
the relevant temperatures to the thermal gradients in the sample.
To overcome this issue, we use a relation shown previously by \citep{bahari2018directional}
using COMSOL simulations, which states, for most cases, $dT_{c/h}=0.7\cdot\Delta T_{c/h}$.
}

\subsection{Sample preparation\label{subsec:Sample-preparation}}

\paragraph{Sandwich sample}

We placed a rectangular coverslip (22 x 40 mm, thickness 0.13-0.17
mm SPIsupplies \#01022-AB) on the copper blocks. We then placed a
$25\,\mathrm{\mu L}$ double distilled water droplet in the middle
of the coverslip and covered it with a square coverslip (22 x 22 mm,
thickness 0.13-0.17 mm SPIsupplies \#01023-AB).

\paragraph{Capillary sample}

We placed the an identical rectangular coverslip (SPIsupplies \#01022-AB)
on the copper blocks. We then cut capillaries (VitroTubes TM \#5003,
\#5005, \#5010 with heights of 30, 50, 100{\microns} and widths
of 300, 500, 1000{\microns} on the inner channels, respectively)
with diamond scratch pen to a desirable length in order to fit them
into the apparatus. After this, we filled the capillaries with double
distilled water and placed them on the coverslip.

\subsection{Freezing procedure}

During this procedure, we placed an additional droplet of double distilled
water on a rectangular coverslip (SPIsupplies \#01022-AB) at the cold
side of the stage for ice nucleation. We then reduced the temperatures
of the thermal blocks to $\Delta T_{c}=-10{\normalcolor \mathrm{\,^{o}C}};\,\Delta T_{h}=+1\mathrm{\,^{o}C},$ nucleating
ice by manually touching the droplet with an ice crystal. Subsequently,
we increased the temperatures of the blocks to $\Delta T_{c}=-3\mathrm{\,^{o}C;}\,\Delta T_{h}=+2\,\mathrm{^{o}C}$
(unless stated otherwise) and moved the sample so the capillary or
the water between the coverslips would be over the slit between the
blocks. Finally, we set the sample in motion with velocity $v_{s}$.

\subsection{Imaging of ice-water interface}

The images were acquired with Olympus microscope, using a non immersive objective (4×, NA=0.10).
Bright field 1750 x 1315 {\microns} images were captured on DMK 23U274   camera (The Imaging Source, Germany) every 1 sec.

Image processing was performed using the ImageJ public domain software \cite{imagej}. The ice-water interface was determined by applying threshold to the gray-scale images to produce binary images. The interface appeared as 4 $\pm$ 0.5 {\microns} wide stripe. 
The center of the stripe was manually measured as a position of the ice front.
In Figures \ref{fig:steady state} and \ref{fig:Interface-dynamics} error bars are within the data point markers.

\subsection{Data fitting}

To fit the experimental data we used MATLAB software. Fitted values and precision range was obtained with nonlinear regression curve fitting tool.

\section{Experimental results}

\subsection{Moving stage steady state\label{subsec:Moving-stage-steady}}

\begin{figure}
\begin{centering}
\includegraphics[width=12cm]{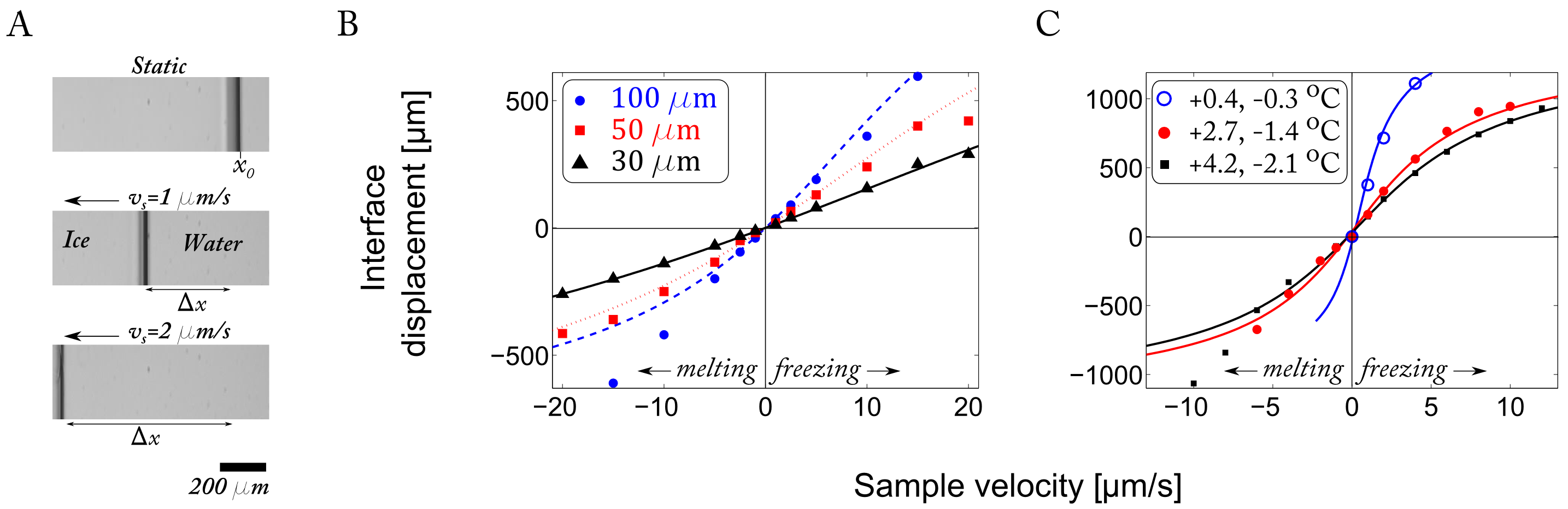}
\par\end{centering}
\centering{}\caption{\label{fig:steady state} Steady-state maximal interface displacement
\emph{$\Delta x$} for different sample velocities $v_{s}$. (A) Snapshots
from an experiment show the interface position when the sample is
static (top panel) and the steady-state position when the sample is
moving toward the cold side with velocities of $1\,\frac{\mathrm{\mu m}}{s}$
(middle panel) and $2\,\frac{\mathrm{\mu m}}{s}$ (bottom panel).
The snapshots are from data set $-0.3\mathrm{\,^{o}C},\,+0.4\mathrm{\,^{o}C}$
in panel C. (B) Ice interface displacement vs. translation velocity
for different capillaries (capillary samples) with water, channel
heights $30,50,100${\microns}, and temperatures $dT_{c}=-2.1\mathrm{\,^{o}C},\,dT_{h}=+1.4\mathrm{\,^{o}C}$.
The lines correspond to the best fit of Eq. (\ref{eq:freezing velocity steady state})
with $\eta$ as a free parameter. (C) Ice interface displacement vs.
translation velocity for a water sample between two coverslips (sandwich
samples) for different block temperatures. The lines correspond to
Eq. (\ref{eq:freezing velocity steady state}) and were fitted to
the experimental data with free parameters $dT_{c}$ and $\,dT_{h}$.}
\end{figure}

As the sample moves toward one of the thermal blocks at velocity $v_s$ the ice front moves, i.e. ice grows or melts, with velocity $v_f$ in the opposite direction.
The system reaches steady state when the ice front moves at the speed of the sample ($v_{f}=v_{s}$). 
During translation, the position of the ice front is displaced by $\Delta x$ relative to its position when the sample is static $x_0$

We measured the steady state displacement of the ice front $\Delta x$ relative to its position before the sample was set in motion $x_{0}$  (Fig. \ref{fig:steady state}A). 
We measured $\Delta x$  at different transnational velocities (Fig. \ref{fig:steady state}B,C) for both freezing ($v_{s}>0$) and melting ($v_{s}<0$).
Faster sample translation velocities resulted in larger $\Delta x$ (Fig. \ref{fig:steady state}B,C).
We also observed larger $\Delta x$  in the thicker capillaries (Fig. \ref{fig:steady state}B).

Next we tested how the temperature gradient affects the displacement  using a sandwich sample varying the temperatures of the thermal blocks $dT_{c}$ and $dT_{h}$. 
Here we found that larger temperature differences between the thermal blocks produced smaller displacements, $\Delta x$ (Fig. \ref{fig:steady state}C).

\subsection{Accelerating freezing front (transient phase) \label{subsec:Initiation-of-freezing}}
Immediately with the initiation of sample movement ($t=0$), the ice front accelerated until it reached the sample velocity \cite{rubinsky1985cryomicroscope}. 
During this phase, the relation between the freezing velocity $v_{f}$ and the sample velocity $v_{s}$ can be written as $v_{f}=v_{s}-\frac{d\Delta x}{dt}$.

\begin{figure}
\begin{centering}
\includegraphics[width=9cm]{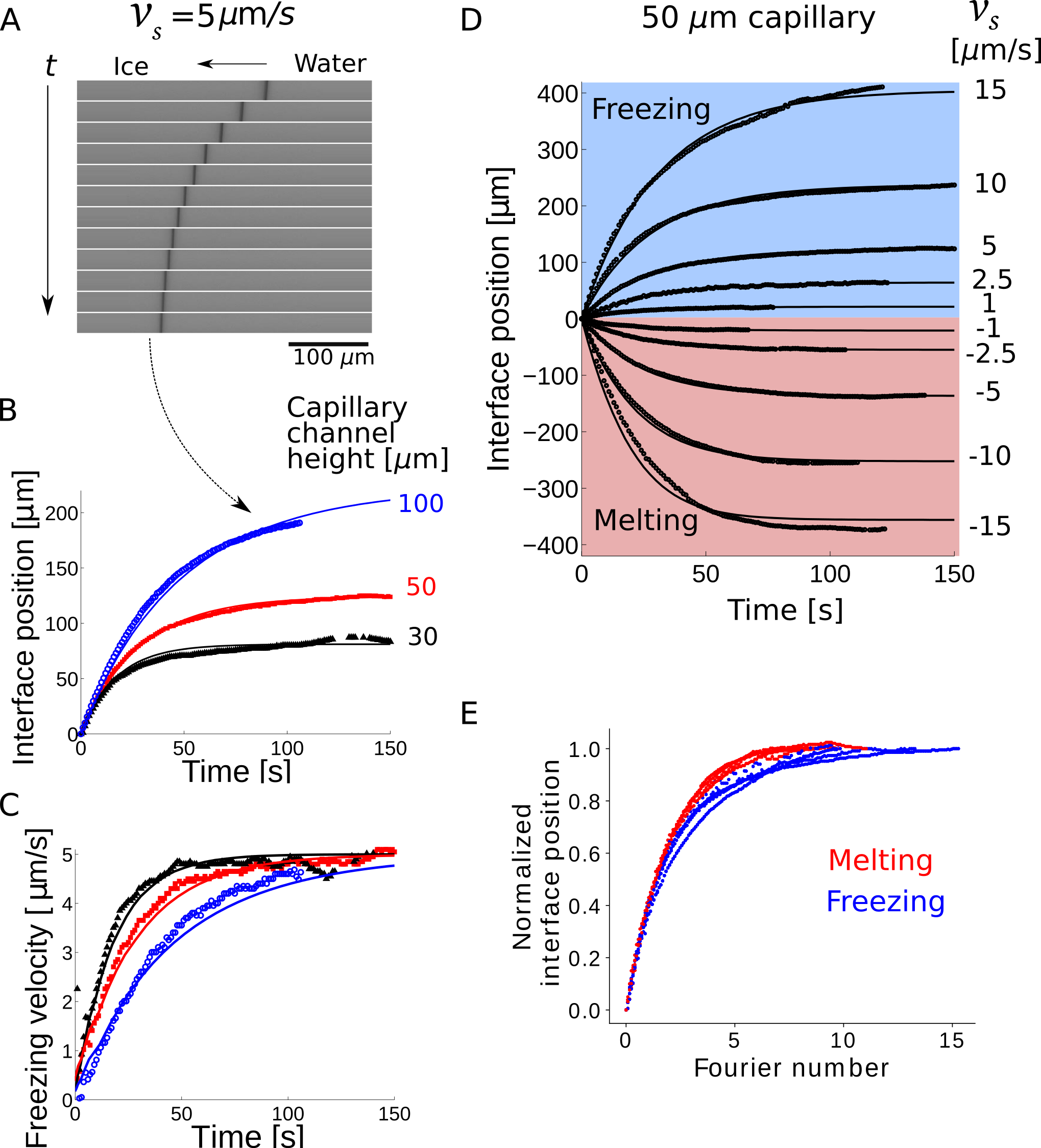}
\par\end{centering}
\caption{\label{fig:Interface-dynamics}Interface position $\Delta x(t)$ during
ice growth acceleration. (A) A sequence of ice front snapshots at
different times after the beginning of motion. (B-C) A comparison
of the dynamics for three different capillaries as a function of time for sample velocity $v_s=5 \frac{\mathrm{\mu m} }{s} $. 
(B) ice front position
$\Delta x(t)$  (C) Freezing velocity ( $v_{f}=v_{s}-\frac{d\Delta x}{dt}$).
(D) ice front position $\Delta x(t)$ for sample velocities $v_{s}$ in the range of -15
to 15 $\frac{\mathrm{\mu m}}{s}$ for a $50${\microns} capillary.
(E) Normalized  ice front position $\frac{\Delta x(t)}{\Delta x}$ as a function of dimensionless time, Fourier number
(same data as in panel D, without $\pm$1 $\frac{\mathrm{\mu m}}{s}$).
Each solid line curve in panels B, C, and D is a numerical solution of Eq. (\ref{eq:x diff eq interface dynamics}).}
\end{figure}

We subsequently explored the dynamics of an accelerating ice front upon an instant change in the sample velocity. 
We followed the ice front position as in Section \ref{subsec:Moving-stage-steady}, but here we followed the ice front immediately after the initiation of sample motion until the ice front reached its steady-state position (Fig. \ref{fig:Interface-dynamics}A).
We performed experiments using rectangular capillaries of different heights  $h=30,\,50,\:100${\microns} and plotted the position of the ice front as a function of time (Fig. \ref{fig:Interface-dynamics}B).

As we had previously observed (Fig. \ref{fig:steady state}B), the displacement of the ice front in thicker capillaries was larger (Fig. \ref{fig:Interface-dynamics}B). 
The velocity of the ice front $v_{f}$ started from static $v_{f}=0$ and accelerated to $v_{f}=v_{s}$ (Fig. \ref{fig:Interface-dynamics}C). 
In the imaging frame, the ice front velocity $\frac{d\Delta x}{dt}$ started from $v_{s}$, meaning that the ice front was not moving, and went down to 0, when the ice front velocity reached the sample velocity, i.e. steady-state.
In thinner capillaries, ice front reached steady-state velocity faster and the dynamics are qualitatively different between the capillaries (Fig. \ref{fig:Interface-dynamics}C).

Next, we tested the effect of the sample velocity on the ice front dynamics in the range $v_s=1-15\,\frac{\mathrm{\mu m}}{s}$ (showing the results for the 50 {\microns} capillary in Fig. \ref{fig:Interface-dynamics}D).
Higher velocities resulted in larger displacement of the ice front, similar to (Fig. \ref{fig:steady state}B).
To compare the dynamics of the ice front at different sample velocities, we normalized the ice front position by its steady-state displacement (Fig. \ref{fig:Interface-dynamics}E), and by changing
the time to be dimensionless Fourier number $F=\frac{\alpha t}{L^2}$ , where $\alpha =0.5 ~\frac{\mathrm{mm}^2}{s}$ is the mean thermal diffusivity of water and ice, and $L=2.6 ~\mathrm{mm}$ is the distance between the thermal blocks. 
The normalized ice front positions collapsed into two trajectories that correspond to melting and freezing.

\section{Theory}
The position of the ice front is set by the energy conservation of the heat flux at the ice front.
Heat flow is approximately one-dimensional in the sample (\ref{sec:Is-heat-flux-horizontal}).

\subsection{Static sample}
Heat flows through the liquid and the glass in the sample from the hot side, at temperature $T_{m}+dT_{h}$, to the cold side, at temperature $T_{m}-dT_{c}$.
According to Fourier's law, the heat flux is $J=kG$, where \emph{k} is the thermal conductivity and \emph{G} is the temperature gradient. 
Energy conservation requires the heat flux in the liquid on the hot side of the ice front equals the heat flux through the ice from the interface on the cold side
\[
k'_{i}G_{i}=k'_{w}G_{w}\:,
\]
where $G_{i}/G_{w}$ are the temperature gradients in the ice/water
and $k'_{i}/k'_{w}$ are the effective thermal conductivities in the
ice/water, which makes $k'_{i/w}=\eta k_{ice/water}+(1-\eta)k_{glass}$.
$\eta$ the ratio of the cross-section of the medium (ice/water) relative
to the whole cross-section that heat flows through, including the
container (glass), and goes between 0 (heat flowing only through the
container) to 1 (heat flowing only through the freezing medium). 
The gradients are $G_{i}=\frac{dT_{c}}{x_{0}}$ and $G_{w}=\frac{dT_{h}}{L-x_{0}}$,
where $x=x_{0}$ is the ice front position (Fig. \ref{fig:stage-setup}B).

The ice front position is different
for differing capillaries (Fig. \ref{fig:stage-setup}D), implying that the thermal conductivity and geometry of the container in which the liquid water is placed
affect the heat flow and must be considered when calculating the ice front position. 
This dependence is discussed in Section (\ref{subsec:Effect-of-container}) along with the meaning and calculation of $\eta$.

\subsection{Steady-state constant freezing velocity}

The displacement of the ice front relative to its static position, $\Delta x$, stems from the latent heat generated during ice growth.
The latent heat generation rate is given by $\eta\rho\Delta H_{f}v_{f}$ , where $\Delta H_{f}$ is the latent heat per unit mass and $\rho$ is the density of ice at the freezing point.

The conservation of the heat flux at the interface is expressed by the Stefan condition, assuming constant gradients (quasi-static approximation) which gives 

\begin{eqnarray}
\eta\rho\Delta H_{f}v_{f}+k'_{w}G_{w} & = & k'_{i}G_{i}\;.\label{eq:Appendix-heat flux}
\end{eqnarray}
Inserting the expressions for the gradients and the effective thermal
conductivities for slow freezing velocity assuming constant thermal
gradients (Sec. \ref{subsec:When-is-ConstGrad}) results in 
\begin{equation}
v_{f}=\frac{\eta k_{ice}+(1-\eta)k_{glass}}{\eta\rho\Delta H_{f}}\frac{dT_{c}}{x_{0}-\Delta x}-\frac{\eta k_{water}+(1-\eta)k_{glass}}{\eta\rho\Delta H_{f}}\frac{dT_{h}}{L-x_{0}+\Delta x}\;,\label{eq:freezing velocity steady state}
\end{equation}

which relates the displacement of the ice front $\Delta x$ to the
translational velocity of the sample $v_{s}=v_{f}$ at steady-state.
We examine this relation, $\Delta x(v_{s})$, experimentally in Section
\ref{subsec:Moving-stage-steady} (Fig. \ref{fig:steady state}).

\subsection{Transient-phase accelerating freezing velocity}

During the acceleration, the relation between the freezing velocity
$v_{f}$ and the sample velocity $v_{s}$ can be written as $v_{f}=v_{s}-\frac{d\Delta x}{dt}$.
Inserting this expression for $v_{f}$ to Eq. (\ref{eq:freezing velocity steady state})
yields 
\begin{equation}
v_{s}-\frac{d\Delta x}{dt}=\frac{k'_{i}}{\eta\rho\Delta H_{f}}\frac{dT_{c}}{x_{0}-\Delta x}-\frac{k'_{w}}{\eta\rho\Delta H_{f}}\frac{dT_{h}}{L-x_{0}+\Delta x}\;,\label{eq:x diff eq interface dynamics}
\end{equation}
which is a differential equation on $\Delta x(t)$ with the initial
condition $\Delta x(0)=0$. This treatment can be applied to the instant change of the velocity. We solve this equation
numerically, and test the solution experimentally in Section \ref{subsec:Initiation-of-freezing}
(Fig. \ref{fig:Interface-dynamics}).

\section{Comparison of experimental results to theory}

\begin{table}
\centering
\begin{tabular}{|c|c|c|c|c|}
\hline 
\multirow{2}{*}{Capillary height} & \multicolumn{2}{c|}{Fitted - $\eta$} & \multicolumn{2}{c|}{Calculated - $\eta$}\tabularnewline
\cline{2-5} \cline{3-5} \cline{4-5} \cline{5-5} 
 & Value & Range & Value & Range\tabularnewline
\hline 
\hline 
30{\microns}  & 0.11 & 0.10-0.12 & 0.13 & 0.12-0.15\tabularnewline
\hline 
50{\microns} & 0.20 & 0.17-0.22 & 0.19 & 0.18-0.21\tabularnewline
\hline 
100{\microns} & 0.35 & 0.23-0.47 & 0.29 & 0.27-0.31\tabularnewline
\hline 
\end{tabular}

\caption{Values and confidence range of $\eta$ obtained by fitting the ice-water interface position shown in figure \ref{fig:steady state}B with  Eq. (\ref{eq:freezing velocity steady state}) and calculated from the sample geometry accordingly.}

\label{table:eta_fitted}
\end{table}


For both the steady-state (Fig. \ref{fig:steady state}) and the transient phase (Fig. \ref{fig:Interface-dynamics}) we fit our model to the experimental results.
We first fit the  model to the experimental results for the interface displacement as a function of sample velocity in steady-state ice growth to Eq. (\ref{eq:freezing velocity steady state}).
In the first data set we used different capillaries (Fig. \ref{fig:steady state}B), using $\eta$ as a fitting parameter.
The fitted values of $\eta$ are compared to its calculated value (see Sec. \ref{subsec:Effect-of-container} for explanation of calculation), which were estimated from the physical dimensions of the sample, see Table~\ref{table:eta_fitted}.
The thin wall approximation (Sec. \ref{subsec:Effect-of-container}) was valid for the 30 and 50{\microns} capillaries, but was less accurate for the larger 100{\microns} capillary.

The displacement ($\Delta x$) increased when
the freezing medium occupied a larger cross-section, namely larger $\eta$. Two intuitive limits of $v\rightarrow0$ and $\eta\rightarrow0$
resulted in no shift in the position of the interface $\Delta x\rightarrow0$,
because either no latent heat was generated or the generated latent
heat was negligible relative to the heat flux generated by the thermal
gradient.

In the second data set we varied the temperatures of the thermal blocks (Fig. \ref{fig:steady state}C) fitting the data by Eq. (\ref{eq:freezing velocity steady state}) with $dT_h$ and $dT_c$ as the fitting parameters.
The results of the fitting of the temperatures were found to be consistent with experimental values of the temperatures on the edges of the thermal blocks themselves, see Table~\ref{table:dT_h-c-fitted}.

As the temperature difference between the thermal blocks decreased, so did the thermal gradients in water and ice. 
The ice front displacement increased to keep the difference between the thermal gradients in water and ice sufficient to evacuate the latent heat from the ice front (Fig. \ref{fig:steady state}C).

The uncertainty in the thermal gradient becomes very large when the  ice front is close to one of the thermal blocks. This could explain the  deviation between the experimental data and the model   for a large displacement in either freezing or  melting (\ref{sec:Interface-position_validity}).


\begin{table}
\centering
\begin{tabular}{|c|c|c|c|}
\hline 
\multicolumn{2}{|c|}{Experimental Temperatures} & \multicolumn{2}{c|}{Fitted Temperatures}\tabularnewline
\hline 
\multicolumn{1}{|c|}{$dT_{h}\,(^{\circ}C)$} & \multicolumn{1}{c|}{$dT_{c}\,(^{\circ}C)$} & \multicolumn{1}{c|}{$dT_{h}\,(^{\circ}C)$} & \multicolumn{1}{c|}{$dT_{c}\,(^{\circ}C)$}\tabularnewline
\hline 
\hline 
+0.40$\pm$0.02 & -0.30$\pm$0.02 & +0.65$\pm$0.01 & -0.42$\pm$0.01\tabularnewline
\hline 
+2.70$\pm$0.02 & -1.40$\pm$0.02 & +2.05$\pm$0.16 & -0.78$\pm$0.12\tabularnewline
\hline 
+4.20$\pm$0.02 & -2.10$\pm$0.02 & +2.4$\pm$0.3 & -0.66$\pm$0.14\tabularnewline
\hline 
\end{tabular}

\caption{Temperatures on the edges of the thermal blocks, comparison of experimental
values to the fitted values in figure \ref{fig:steady state}C.}
\label{table:dT_h-c-fitted}

\end{table}

In the transient phase, the experimental results of the ice front displacement vs. sample velocity were fitted by Eq. (\ref{eq:x diff eq interface dynamics}).
The experimental data was found to be in agreement with the numerical solution. For 28 experiments  $R^2$ was found to be above 0.97 and for $v_s=1  ~\frac{\mathrm{\mu m}}{s}$ for the 30 {\microns} capillary and  $v_s=-15  ~\frac{\mathrm{\mu m}}{s}$ for the 100 {\microns} capillary its values were  $R^2=0.94,0.95$ respectively. 
The fitted values are $\eta$=0.08-0.13, 0.15-0.21, 0.29-0.40 for capillary heights of h=30, 50, 100 \microns, respectively.

\section{Discussion}

\subsection{Practical experimental considerations}

\subsubsection{The effect of the container thickness and the meaning of $\eta$\label{subsec:Effect-of-container}}

Typical freezing experiments are executed when liquid is placed inside
a container, such as between two glass coverslips or within a capillary.
Herein, we address the practical aspects, assuming a glass-made container.
Heat generated at an ice-water interface is removed through the sample,
which contains the ice and the container, toward the cold block. Assuming
a unidirectional heat flux between the thermal blocks, the effective
thermal conductivities in the ice and the water sides of the sample, are given by $k'_{i}(\eta)$ and $k'_w(\eta)$ that we defined above, respectively.

If the container walls are very thin, heat flows through the whole cross-section of the sample $A_{total}$. 
In this case, $\eta=\frac{A_{l}}{A_{total}}$, where $A_{l}$ is the cross-section of the liquid. However,
in cases where the cross-section of the container is very large relative
to that of the liquid fraction, only the part of the container that
is close to the liquid fraction contributes to the thermal conduction
of the latent heat generated at the ice-water interface. In such cases,
$\eta$ must be larger than its geometric value to
describe the effective thermal conductivities $k'_i$ and $k'_w$.

\subsubsection{The approximation of constant thermal gradients  \label{subsec:When-is-ConstGrad}}

The heat generated by the moving freezing front is transferred by
both diffusion (through conduction) and convection through the uniform
displacement of the sample with velocity $v_{s}$ in thermal gradient
through a typical distance $L$. The Peclet number $Pe=\frac{v_{s}L}{\alpha}$,
where $\alpha$ is the thermal diffusivity, represents the ratio between
heat convection and heat diffusion through conduction in the system.
When $Pe\ll1$, heat transport is dominated by conductivity, thus,
the thermal gradients could be considered constant (\ref{sec:Sample-velocity-Pe}).
For a water-ice sample, this condition is satisfied for translation
velocities $v_{s}<20\,\frac{\mathrm{\mu m}}{s}$. Notably, most experiments
using the directional stage in cryopreservation \citep{rubinsky1985cryomicroscope},
ice lens growth \citep{schollick2016segregated,anderson2012periodic},
and emulsion freezing \citep{dedovets2018five}, are realized within 
this velocity range.

\subsection{Kinetic attachment coefficient of ice crystal growth}

The kinetic attachment coefficient $\beta$ determines the dependence of the ice growth
velocity's on the degree of supercooling in the ice-water
interface $\Delta T_{SC}$ by $v_{f}=\beta\cdot\Delta T_{SC}$ \citep{libbrecht2017physical}.
Therefore, some of the interface displacement $\Delta x$ in the translational
gradient stage during freezing and melting can be attributed to the
interface supercooling.

To estimate this effect, we have used a value of $\beta=2\cdot10^{-4}\frac{m}{s\cdot K}$ \citep{libbrecht2017physical}. 
Our estimation shows the effect of interface supercooling  on $\eta$ within the error range of our fitting.
Our model neglects this effect.

\section{Summary}

We defined the freezing and melting steady-states as states in which the
ice front moved at a constant velocity $v_{f}$ equal to
the sample translation velocity $v_{s}$. In steady-state, the heat
flux balance calculation (Eq. (\ref{eq:freezing velocity steady state}))
predicted a shift of the interface position $\Delta x$ relative to
its static position $x_{0}$ as a function of $v_{s}$. Our experiments
confirmed this prediction. We measured $\Delta x$ in a range of translational
velocities $v_{s}$ and found $\Delta x(v_{s})$ in good agreement
with the calculated values (Eq. (\ref{eq:freezing velocity steady state})).
Quantitative predictions of $\Delta x$ required considering
the specific geometry of each sample, specifically the ratio of the cross sections of the liquid and the container, $\eta$, and the boundary temperatures $T_{m}-dT_{c}$ and $T_{m}+dT_{h}$.
A comparison of experimental values of $\eta$, $dT_{c}$, and $dT_{h}$
with calculated values showed agreement, as long as two conditions
justifying the constant gradient approximations were met: First, the
sample velocity was kept small, i.e., $Pe<0.1$, (\ref{sec:Sample-velocity-Pe});
Second, the interface position was not close to a thermal block edge
(\ref{sec:Interface-position_validity}).

Next, we explored the transient phase \citep{alexandrov2007nonlinear,javierre2006comparison} between two steady states upon sudden velocity change.
We found that there is a typical time scale in which the ice front position shifted towards its new steady-state position. 
The dynamics of the shift $\Delta x(t)/\Delta x$ (normalized to the steady-state value) were found to be independent of the freezing or melting velocity $v_{s}$ (Fig. \ref{fig:Interface-dynamics}E).
This result is non-trivial, since it does not appear as a direct conclusion
from Eq. (\ref{eq:x diff eq interface dynamics}).
Our results show that samples with smaller $\eta$ values reached new steady-states faster (Fig. \ref{fig:Interface-dynamics}C).

Characterization and understanding of the transient phase are relevant for systems in which ice grows with a non-constant velocity, such as during ice lens growth. 
The observed ice lens growth, namely the acceleration of ice growth, is similar to the ice-water interface dynamics reported here (Fig. \ref{fig:Interface-dynamics}) \citep{schollick2016segregated,wang2016interface}.
Our analysis referred to pure substance.  
The heat flux balance approach that we describe should be part of the analysis also when a system that includes solutes is solved. 

Overall, this work has provided an experimentally validated theoretical framework for the  experiment design of translational freezing setup. 
For example, Eq. (\ref{eq:freezing velocity steady state}) allows the determination the interface position for a specific freezing velocity in the imaging field by adjusting the temperatures of the thermal blocks in the steady-state regime. 
While, in a transient regime, one can control the duration of the transient phase through a sample design and container geometry.

\section*{Acknowledgements}

IB acknowledges support from the Israel Science Foundation grant number
930/16. MC acknowledges support from The Samuel and Lottie Rudin Scholarship
Foundation.

\appendix

\section{The sample velocity $v_{s}$ effect on the quasi-static(constant gradient) approximation\label{sec:Sample-velocity-Pe}}

When a sample is in motion, heat is transferred by convection in addition
to conduction with the corresponding heat equation 
\begin{equation}
\frac{\partial T_{n}}{\partial t}=\alpha_{n}\frac{\partial^{2}T_{n}}{\partial x^{2}}+v_{s}\frac{\partial T_{n}}{\partial x}\:,
\end{equation}
where $n=i,w$ represents ice and water, respectively, and $\alpha$
is the thermal diffusivity. We are interested in a steady-state solution,
i.e., $\frac{\partial T}{\partial t}=0$. 
In such case the heat equation is reduced to $\alpha_{n}\frac{\partial^{2}T_{n}}{\partial x^{2}}=-v_{s}\frac{\partial T_{n}}{\partial x}$, that means that the correction to the thermal gradient, i.e. $\frac{\partial^{2}T_{n}}{\partial x^{2}}$, is proportional to the gradient itself with the proportion being  $\frac{v_s}{\alpha}$.
The distance $\frac{v_s}{\alpha}$ should be compared to the typical distance of thermal profile \emph{L}.
If $Pe=\frac{v_{s}L}{\alpha_{n}}\ll1$,
the gradients in water and ice, $G_{w}$ and $G_{i}$,
respectively, can be considered constant. For typical experimental conditions (i.e., $\alpha_{i}=0.84\frac{mm^{2}}{s},\,\alpha_{w}=0.13\frac{mm^{2}}{s}$ \citep{james1968thermal} and $L=2.6$ mm), the constant gradients approximation is valid for sample velocities $v_{s}<20\frac{\mu m}{s}$.

For the transient phase this justification that is based on steady-state is not valid.
Then, we look at the time scale to establish thermal gradient upon change in interface position $L^2/\alpha$ which is on the order of a second, while the time scale to reach new steady-state position $L/v_s$ is on the order of 100 seconds.
The ratio of the time scales is the Peclet number.
When the Peclet number is small, the time scale to establish constant thermal gradients is much smaller than the other time scale in the system.
So, we can separate the time scales, and treat the thermal gradient as constant at the time scale of the experiments.

In the opposite limit (i.e., a fast freezing regime), the full heat
equation $\alpha_{n}\frac{\partial^{2}T_{n}}{\partial x^{2}}+v_{s}\frac{\partial T_{n}}{\partial x}=0$
must be solved to obtain the temperature profile \citep{rabin2000reasonable,lipp1987investigation}.

Note that using a gap size \emph{L} for calculating \emph{Pe} gives
an overestimate, since only the distance between the interface to
each side is relevant.

\section{The interface position between the thermal blocks effect on the constant
gradient approximation\label{sec:Interface-position_validity}}

The temperature profile near the edge of a thermal block changes smoothly
from a linear profile, which is equal to the thermal gradient \emph{G,}
to a constant temperature, which is equal to the temperature of the
block. If the interface is close to this transition zone, then the
temperature profile can not be approximated by a linear profile.

For example, in the 100{\microns} capillary data of Fig. \ref{fig:steady state}B
and in the melting velocities above 10 $\frac{\mathrm{\mu m}}{s}$
of Fig. \ref{fig:steady state}C, there are discrepancies between
the data and the fit to Eq. (\ref{eq:freezing velocity steady state}).
These discrepancies are due to the interface being relatively close
to the hot thermal block (e.g., $L-x_{0}+\Delta x$), making the expression
$\frac{dT_{h}}{L-x_{0}+\Delta x}$ in Eq. (\ref{eq:freezing velocity steady state})
sensitive to an uncertainty in $L$, which also comes from the temperature
profile near the edges of the blocks.

\section{Is the heat flux horizontal and where? \label{sec:Is-heat-flux-horizontal}}
\begin{figure}
\begin{centering}
\includegraphics[width=10cm]{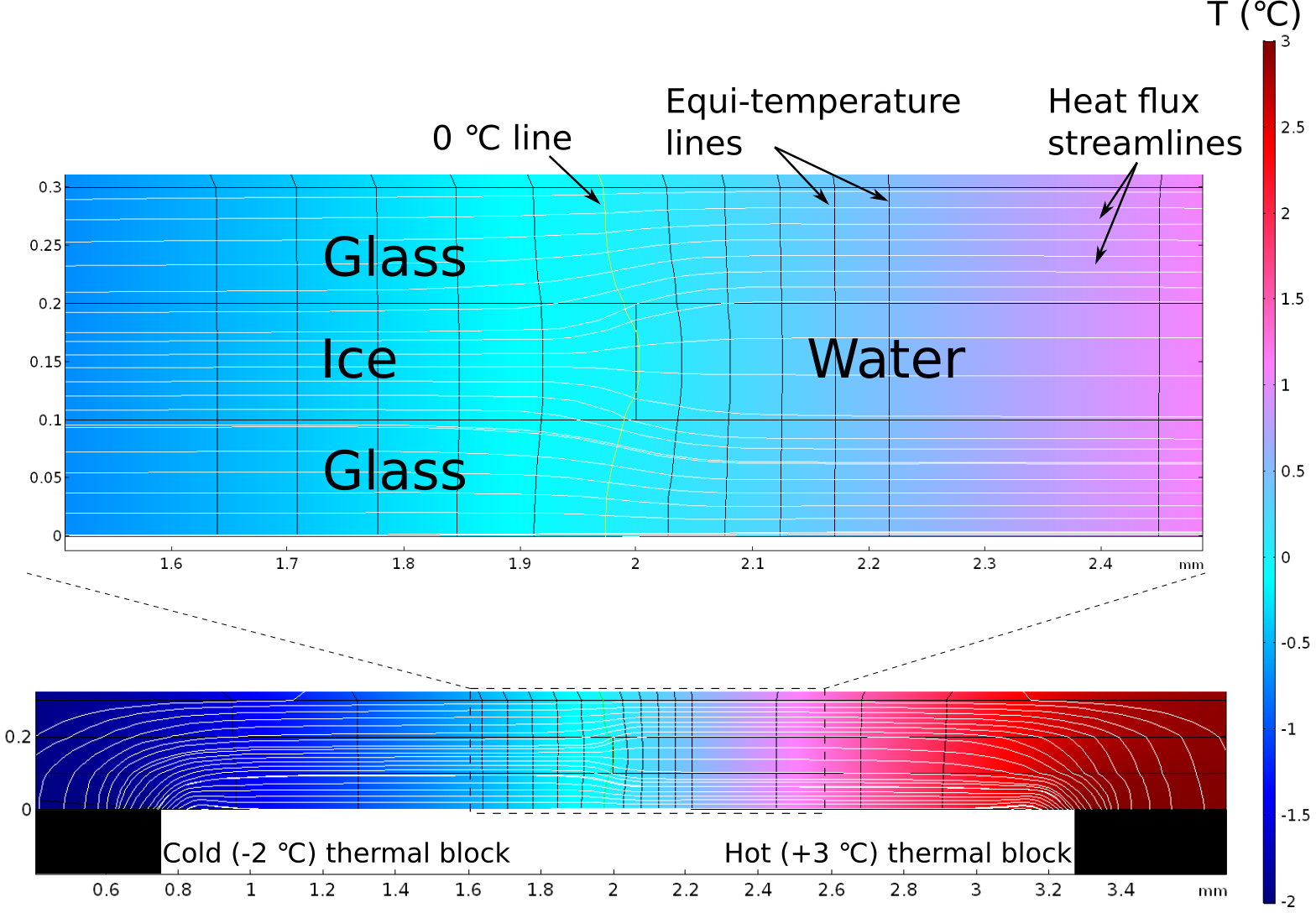}
\par\end{centering}
\caption{\label{fig:heat-flux-simulation}Finite element solution of the temperature distribution (color mapped and equi-temperature black lines) and the heat flux (white streamlines) in freezing water sample inside glass container placed between two thermal blocks. Lower panel shows the result of the simulation featuring the sample of glass-water/ice-glass and the two thermal blocks that were simulated as a fixed temperature constraints on the edges of the sample as is drawn. Upper panel shows a zoom of  the sample at the ice front.  Comsol solver was used to solve the heat equation.
}
\end{figure}

In the model we assume that the heat flux in the sample from the hot to the cold side is uni-directional.
To test this approximation we performed a simulation of the temperature distribution and heat flow in the sample.
The simulation performed in 2d of the x-z cross section of the sample, while in the y direction the sample is infinite (Fig. \ref{fig:heat-flux-simulation}).
We find that the heat flux is horizontal and uni-directional throughout the sample except on the two thermal blocks edges and the ice-water interface.
Near the edges there is a region from the thermal block into the gap between the blocks (where the imaging takes place) of $\sim$200 \microns \, length in which there is a vertical heat flux that is caused by having the thermal blocks below the glass.
Second, there is a vertical heat flux around the ice front ($\sim$100 \microns) because of the transition of the heat flow between the two different regimes on the ice and the water sides.
On the water side of the sample, heat flows are horizontal through water and glass.
Since, thermal conductivity of glass is higher than that of water, the heat flux in the water part of the sample concentrates within the glass and the density of the heat streamlines there is larger than in water (Fig. \ref{fig:heat-flux-simulation}).
Oppositely, thermal conductivity of ice is higher than glass, so the heat flow in the frozen part of the sample concentrates within the ice (Fig. \ref{fig:heat-flux-simulation}).
Therefore, at the interface between water and ice the heat flux has a vertical component (upper panel, Fig. \ref{fig:heat-flux-simulation}).

In the simulation the sample is static, which is a valid approximation in case of the slow sample velocity limit, namely when the quasi-static approximation is valid ($Pe\ll1$, see \ref{sec:Sample-velocity-Pe}).
At this limit the temperature distribution is qualitatively the same as when the sample is static, since the convection term can be neglected. 

\bibliographystyle{unsrtnat}
\bibliography{dirctional}

\end{document}